\documentclass[prl,aps,amssymb,amsmath,superscriptaddress,twocolumn,showpacs]{revtex4}

\usepackage{graphicx}

\renewcommand{\epsilon}{\varepsilon}

\newcommand{\kk}{\mathbf{k}}

\newcommand{\p}{\mathbf{p}}
\renewcommand{\phi}{\varphi}
\newcommand{\R}{\mathbb{R}}

\newcommand{\x}{\mathbf{x}}
\newcommand{\Z}{\mathbb{Z}}

\DeclareMathOperator{\tr}{Tr}

\begin{document}

\title{Extended quantum conditional entropy and quantum uncertainty inequalities}
\thanks{\copyright\, 2012 by the authors. This paper may be reproduced, in its entirety, for non-commercial purposes.}

\author{Rupert L. Frank}
\affiliation{Department of Mathematics, Princeton University, Princeton, NJ 08544, USA}

\author{Elliott H. Lieb}
\affiliation{Department of Mathematics, Princeton University, Princeton, NJ 08544, USA}
\affiliation{Department of Physics, Princeton University, P.~O.~Box 708, Princeton, NJ 08542, USA}

\begin{abstract}
 Quantum states can be subjected to classical measurements, whose incompatibility, or uncertainty, can be quantified by a comparison of certain entropies. There is a long history of such entropy inequalities between position and momentum. Recently these inequalities have been generalized to the tensor product of several Hilbert spaces and we show here how their derivations can be shortened to a few lines and how they can be generalized. All the recently derived uncertainty relations utilize the strong subadditivity (SSA) theorem; our contribution relies on directly utilizing the proof technique of the original derivation of SSA.
\end{abstract}


\maketitle


A celebrated inequality of Maassen--Uffink \cite{MaUf}, based on earlier work in \cite{Hi,De,Kr}, relates the `classical' entropy of a quantum density matrix in two different bases and shows that although one of them could be very small, the sum of the two is bounded below by a positive constant. The word `classical' refers to $H=-\sum_j p_j \ln p_j$, where $p_j$ is the expectation of a density matrix in some orthonormal basis. The inequality is
$$
H(A)+H(B)\geq -2\ln \sup_{j,k} |\langle a_j,b_k\rangle| \,,
$$
where $A$ and $B$ represent two orthonormal bases $(a_j)$ and $(b_k)$. This can be generalized to continuous bases, like position $\x$ and momentum $\p$, in which case the inequality becomes (with $\p=2\pi\kk$)
$$
-\int \!d^d\x\, \rho(\x,\x)\ln\rho(\x,\x) - \int \!d^d\kk\, \widehat\rho(\kk,\kk)\ln\widehat\rho(\kk,\kk) \geq 0 \,.
$$
These inequalities have subsequently been improved; see the review \cite{WeWi} and the recent papers \cite{CoYuGhGr,FrLi,Ru}. Only purely classical analysis needs to be utilized to prove these inequalities and only one Hilbert space is involved.

Another direction was opened up by the conjectures of Renes and Boileau \cite{ReBo}, in which more than one Hilbert space appears. The analogous inequalities become more difficult mathematically because of the well-known entanglement problems in quantum mechanics. Indeed, as noted in \cite{ReBo}, the Lieb--Ruskai strong subadditivity (SSA) theorem \cite{LiRu}, or one of its equivalents, would ultimately be needed to prove the conjectures. Berta \emph{et al.} \cite{BeChCoReRe} then proved a special 2--space version of the conjecture, that can be called the rank-one version, and used \cite{LiRu} again (this time the concavity of conditional entropy, which is equivalent to SSA) to deduce a 3--space version of the uncertainty principle. This quantum version of the uncertainty principle has attracted a great deal of attention.

Subsequently, Coles \emph{et al.} \cite{CoYuGhGr} and Tomamichel and Renner \cite{ToRe} were able to eliminate the rank one condition for the 3--space version and partially removed it for the 2--space version. The significant developments in \cite{BeChCoReRe,CoYuGhGr} came at the cost of rather lengthy calculations and multi-page proofs.

It is evidently desirable to shorten these proofs and to clarify the essential mathematical underpinning. It is shown here that if one uses the original proof structure of SSA in \cite{LiRu} these proofs can be significantly shortened.

A second thing we do in this paper is eliminate the aforementioned restriction of \cite{CoYuGhGr} in the 2--space case. Those authors point out that the obvious extension is necessarily false; we find a more general formulation that makes the extension possible.

Finally, we go back to the Maassen--Uffink inequality above and relate the $\x$-space entropy to the $\p$-space entropy, but this time with an auxiliary quantum system, thereby promoting the Maassen--Uffink inequality to a truly quantum one.

The notation in this field is not uniform and we begin by defining our terms. We have three Hilbert spaces (degrees of freedom) $\mathcal H_1$, $\mathcal H_2$ and $\mathcal H_3$ and their tensor products $\mathcal H_{123}= \mathcal H_1 \otimes \mathcal H_2 \otimes \mathcal H_3$, $\mathcal H_{12}= \mathcal H_1 \otimes \mathcal H_2$ etc. We have density matrices $\rho_{123}$, $\rho_{12}$, etc. (i.e., non-negative operators of trace one). If $\rho_{123}$ is defined on $\mathcal H_{123}$ then there is a natural $\rho_{12}$ on $\mathcal H_{12}$ given by $\rho_{12} = \tr_3 \rho_{123}$, where $\tr_3$ is the partial trace over $\mathcal H_3$. In general, entropy is defined by $S(\rho) = -\tr_{\mathcal H} \rho\ln\rho$, where $\mathcal H$ is one of the above spaces. If $\rho$ and $\mathcal H$ are understood, then $S_{12} = S(\rho_{12})$ and $S_1=S(\rho_1)$. Conditional entropy is defined by
$$
S(1|2) = S_{12}-S_2 = S(\rho_{12}) - S(\rho_2)
$$
and, as shown in \cite{LiRu}, this is a concave function of $\rho_{12}$. This concavity is mathematically equivalent to SSA, and we use it in the Lemma below.

A classical measurement on $\mathcal H_1$ is defined by a sequence of operators $A_1$, $A_2$, etc. on $\mathcal H_1$ such that $\sum_j A_j^* A_j=1$. We will need two measurements in our discussion, so we will need a second sequence $B_1$, $B_2$, etc. on $\mathcal H_1$ such that $\sum_k B_k^* B_k =1$. At the end of this paper we will let the indices $j$ and $k$ be continuous and the sums become integrals, but for simplicity we stay with sums for now. A measurement $A$ is called `rank-one' if every $A_j$ is a rank-one matrix, namely, $A_j= |a_j\rangle\langle a_j|$.

The classical/quantum conditional entropy corresponding to $A$ and $\rho$ is
\begin{align*}
H(1^A|2) = & -\sum_j \tr_2 \left(\tr_1 A_j\rho_{12} A_j^*\right) \ln \left(\tr_1 A_j\rho_{12} A_j^*\right) \\
& -S(\rho_2) \,.
\end{align*}
Note where the summation sits. (This is \emph{not} the conditional entropy of $\sum_j A_j\rho A_j^*$.) The quantity $H(1^B|2)$ is defined similarly. 

\medskip

\textbf{2--space theorem.}--- For any $\rho_{12}$,
\begin{equation}
 \label{eq:2sp}
H(1^A|2) + H(1^B|2) \geq S(1|2) -2\ln c_1 \,,
\end{equation}
where $c_1=\sup_{j,k} \sqrt{ \tr_1 B_k A_j^*A_j B_k^*}$.

\medskip

In the special case that $\rho_{12}=\rho_1\otimes\rho_2$ is a product state one easily sees that \eqref{eq:2sp} reduces to the 1--space theorems in \cite{CoYuGhGr,FrLi,Ru}, which extend the Maassen--Uffink theorem.

Coles \emph{et al.} \cite{CoYuGhGr} prove \eqref{eq:2sp} if either $A$ or $B$ is rank-one and if $c_1$ is replaced by the smaller, better, number
$$
c_\infty = \sup_{j,k} \sqrt{ \|B_k A_j^*A_j B_k^*\|_\infty }
$$
They state correctly that the theorem cannot hold for general $A$ and $B$ in the $c_\infty$ version. However, if either $A$ or $B$ is rank-one, then $c_1=c_\infty$. So our improvement of Coles \emph{et al.} consists of eliminating the rank-one condition and using $c_1$.

The norm $ \|B_k A_j^*A_j B_k^*\|_\infty$ is the largest eigenvalue of $B_k A_j^*A_j B_k^*$, which is also the largest eigenvalue of $\sqrt{B_k^* B_k} A_j^*A_j \sqrt{B_k^* B_k}$, and thus coincides with $\|\sqrt{A_j^*A_j} \sqrt{B_k^* B_k}  \|_\infty^2$. This shows that our $c_\infty^2$ is, indeed, the same as that in \cite[Eq. (57)]{CoYuGhGr}.

\medskip

\textbf{3--space theorem.}\cite{CoYuGhGr}--- For any $\rho_{123}$,
$$
H(1^A|2) + H(1^B|3) \geq -2\ln c_\infty \,.
$$

\medskip

We shall prove the 3--space theorem by first proving the following theorem which is not conveniently expressed in the $H(\cdot|\cdot)$ notation. 

\medskip

\textbf{2$\tfrac12$--space theorem.}--- For any $\rho_{12}$,
$$
H(1^A|2) - \sum_k \tr_{12} B_k\rho B_k^* \ln B_k\rho B_k^* - S(\rho_{12}) 
\geq -2\ln c_\infty \,.
$$

\medskip

This looks like another 2--space theorem but, as we explain below, it is really the 3-space theorem applied to a pure state (i.e., rank-one) $\rho_{123}$. Thus our name.

The key inequality behind the proofs of all our three theorems is the three operator generalization of the Golden--Thompson inequality from \cite{Li}, which was also the key ingredient in the proof of SSA in \cite{LiRu}. It states that for non-negative operators $X,Y,Z$,
\begin{equation}
 \label{eq:gt}
\tr e^{\ln X-\ln Y +\ln Z} \leq \int_0^\infty \!\! dt\, \tr X (Y+t)^{-1} Z(Y+t)^{-1}.
\end{equation}
Note that in the special case $Z=1$ this reduces to the classical Golden--Thompson inequality. This inequality was a byproduct of the proof of concavity of the generalized Wigner--Yanase skew information.

We shall also need the Gibbs variational principle (equivalent to the Peierls--Bogolubov inequality),
\begin{equation}
 \label{eq:gibbs}
\tr \rho h - S(\rho) \geq - \ln\tr e^{-h}
\end{equation}
for any density matrix $\rho$ and any self-adjoint $h$, and the Davis operator Jensen inequality \cite{Da},
\begin{equation}
 \label{eq:choidavis}
\sum_j A_j^* \left(\ln K_j\right) A_j \leq \ln\left( \sum_j A_j^* K_j A_j \right)
\end{equation}
for any positive operators $K_j$ and any $A_j$ with $\sum_j A_j^* A_j =1$.

\medskip

\emph{Proof of the 2--space theorem.} We use \eqref{eq:choidavis} for both $A$ and $B$ to bound
\begin{equation}
 \label{eq:step1}
H(1^A|2) + H(1^B|2) - S(1|2) \geq \tr_{12} \rho_{12}h - S(\rho_{12}) \,,
\end{equation}
with the operator
\begin{align*}
h = \ln\rho_2 & - \ln \sum_j A_j^*A_j \left(\tr_1 A_j\rho_{12} A_j^*\right) \\
& - \ln \sum_k B_k^*B_k \left(\tr_1 B_k\rho_{12} B_k^*\right).
\end{align*}
We do not need to invoke \eqref{eq:choidavis} when $A$ and $B$ are rank-one; in that case \eqref{eq:choidavis} is an \emph{equality} and the proof simplifies further.

Thus, by \eqref{eq:step1} and \eqref{eq:gibbs},
$$
H(1^A|2) + H(1^B|2) - S(1|2) \geq -\ln \tr_{12} e^{-h} \,,
$$
and it remains to show that $\tr_{12} e^{-h} \leq c_1^2$. Now comes the crucial step! We use \eqref{eq:gt} to bound
$$
\tr_{12} e^{-h} \leq  \int_0^\infty dt\, \sum_{j,k} \tr_{12} C_{j,k}(t)
$$
with
\begin{align*}
C_{j,k}(t) & =  A_j^*A_j \left(\tr_1 A_j\rho_{12} A_j^* \right) (\rho_2+t)^{-1} \\ 
& \quad \times B_k^*B_k \left(\tr_1 B_k\rho_{12} B_k^* \right) (\rho_2+t)^{-1} \\
& = A_j^*A_j B_k^*B_k \ D_{j,k}(t) \,.
\end{align*}
Here, $D_{j,k}(t)$ is the operator on $\mathcal H_2$ given by
$$
\left( \tr_1 A_j\rho_{12} A_j^* \right) (\rho_2+t)^{-1} \left( \tr_1 B_k\rho_{12} B_k^* \right) (\rho_2+t)^{-1} .
$$
Thus,
\begin{align*}
\tr_{12} C_{j,k}(t) & = \left( \tr_1 A_j^*A_j B_k^*B_k \right) \tr_2 D_{j,k}(t) \\ 
& \leq c_1^2 \tr_2 D_{j,k}(t) \,.
\end{align*}
We next note that $\sum_j \tr_1 A_j\rho_{12} A_j^* = \sum_j \tr_1 A_j^*A_j\rho_{12}  = \rho_2$, and similarly for the $k$ sum, and obtain
$$
\sum_{j,k} \tr_2 D_{j,k}(t) = \tr_2 \rho_2^2 (\rho_2+t)^{-2} \,.
$$
Thus, since $\int_0^\infty dt\, (\rho_2+t)^{-2} = \rho_2^{-1}$,
$$
\int_0^\infty dt\, \sum_{j,k} \tr_2 D_{j,k}(t) = \tr_2 \rho_2 = 1 \,.
$$
This completes the proof of the 2--space theorem. QED

We shall now show that the 3--space theorem is a corollary of the 2$\frac12$--space theorem, so that the proof of the 2$\frac12$--space theorem will finish everything. We use the following:

\emph{Lemma.}--- $H(1^A|2)$ is a concave function on the set of non-negative operators $\rho_{12}$ on $\mathcal H_{12}$.

\emph{Proof of the Lemma.}---
The idea is to view the sum over $j$ as the trace over an auxiliary space $\mathcal K$ of a matrix that happens to be diagonal in this space, and to apply concavity of the conditional entropy \cite{LiRu} in $\mathcal H_2\otimes \mathcal K$. The details are as follows: Let $(e_j)$ be an orthonormal basis of $\mathcal K$ and consider the operator $\Gamma= \sum_j \left(\tr_1 A_j\rho_{12} A_j^*\right) \otimes |e_j\rangle\langle e_j|$ on $\mathcal H_2\otimes\mathcal K$. As in the proof of the 2--space theorem we have $\Gamma_2=\tr_{\mathcal K}\Gamma=\rho_2$ and therefore
$$
H(1^A|2) = S(\Gamma) - S(\Gamma_2) \,.
$$
This is the conditional entropy of $\Gamma$ with respect to $\mathcal H_2$. Since $\rho_{12}\mapsto\Gamma$ is linear, the asserted concavity follows from the fact that conditional entropy is concave, as shown in \cite[Thm. 1]{LiRu}.
QED

\emph{Proof of the 3--space theorem.}---
It follows from the Lemma that $H(1^A|2) + H(1^B|3)$ is a concave function of $\rho_{123}$. Thus, for the proof we may assume that $\rho_{123}$ is a pure state (rank one). In that case $S(\rho_3)= S(\rho_{12})$ and, since $B_k\rho_{123} B_k^*$ is pure as well, 
$$
H(1^B|3) =  - \sum_k \tr_{12} B_k\rho_{12} B_k^* \ln B_k\rho B_k^* -S(\rho_{12}) \,.
$$
This reduces the inequality of the 3--space theorem to that of the 2$\frac12$--space theorem.
QED

\emph{Proof of the 2$\frac12$--space theorem.}---
The proof runs very parallel to that of the 2--space theorem. Namely, the right side in the desired inequality is bounded from below by $\tr_{12} \rho_{12}\tilde h  - S(\rho_{12})$, where now
\begin{align*}
\tilde h = \ln\rho_2 & - \ln \sum_j A_j^*A_j \left(\tr_1 A_j\rho_{12} A_j^*\right) \\
& - \ln \sum_k B_k^*B_k \rho_{12} B_k^* B_k \,.
\end{align*}
Here we used \eqref{eq:choidavis}. After applying \eqref{eq:gibbs} as before, everything is reduced to showing $\tr_{12} e^{-\tilde h} \leq c_\infty^2$. The crucial ingredient is again \eqref{eq:gt} which now leads to
$$
\tr_{12} e^{-\tilde h} \leq  \int_0^\infty dt\, \sum_{j,k} \tr_{12} \tilde C_{j,k}(t)
$$
with
\begin{align*}
\tilde C_{j,k}(t) & =  A_j^*A_j \left(\tr_1 A_j\rho_{12} A_j^* \right) (\rho_2+t)^{-1} \\ 
& \quad \times B_k^*B_k \rho_{12} B_k^* B_k (\rho_2+t)^{-1} \,.
\end{align*}
At this point the proof diverges somewhat from that of the 2--space theorem. Namely, by cyclicity of the trace we write
$$
\tr_{12} \tilde C_{j,k}(t) = \tr_{12} \left( B_k A_j^*A_j B_k^* \tilde D_{j,k}(t) \right)
$$
with $\tilde D_{j,k}(t)$ given by
$$
\left(\tr_1 A_j\rho_{12} A_j^* \right) (\rho_2+t)^{-1} B_k \rho_{12} B_k^* (\rho_2+t)^{-1} \,.
$$
Since
$$
B_k A_j^*A_j B_k^* \left(\tr_1 A_j\rho_{12} A_j^* \right) \leq c_\infty^2 \left(\tr_1 A_j\rho_{12} A_j^* \right)
$$
and since $(\rho_2+t)^{-1} B_k \rho_{12} B_k^* (\rho_2+t)^{-1} \geq 0$, we have
$$
\tr_{12} \tilde C_{j,k}(t) \leq c_\infty^2 \tr_{12} \tilde D_{j,k}(t) \,.
$$
{}From here, everything is as before:
$$
\sum_{j,k} \tr_{12} \tilde D_{j,k}(t) = \tr_2 \rho_2^2 (\rho_2+t)^{-2}
$$
and
$$
\int_0^\infty dt\, \sum_{j,k} \tr_{12} \tilde D_{j,k}(t) = \tr_2 \rho_2 = 1 \,.
\qquad\qquad\mathrm{QED}
$$

\medskip

At last, we turn to the continuous version, the most
important application being the position-momentum
uncertainty. We start with this case. Take $\mathcal H_1$
to be $L^2(\R^d)$, the square-integrable functions on
$\R^d$. The spaces $\mathcal H_2$ and $\mathcal H_3$ can be
anything. The measurement $A_j^*A_j$ is $|a_j\rangle\langle
a_j|$ where $a_j$ is the delta-function $\delta(\x-\x')$ for
some $\x'$ in $\R^n$. The $j$ becomes $\x'$, which is a
continuous variable, the sum $\sum_j$ becomes the integral
$\int d^d\x'$, and the normalization condition $\sum_j A_j^*
A_j =1$ becomes $\int d^d\x'\, \delta(\x-\x')\delta(\x''-\x')
=\delta(\x-\x'')$. (We realize that the delta-function is not
a function, but all of this can be made rigorous.)
Similarly, the $B_k^*B_k$'s are $|b_k\rangle\langle b_k|$
where the $b_k$ will also not be square-integrable
functions. They will be plane waves $e^{i 2\pi \kk\cdot\x}$
for some $\kk$ in $\R^d$. Again, $\kk$ is a continuous index
and sums are integrals. The normalization condition $\sum_k
B_k^* B_k =1$ becomes $\int d^d\kk \,
e^{-2\pi i \kk\cdot \x} e^{2\pi i \kk\cdot \x'} =\delta(\x-\x')$.

Let $\rho_{12}$ be a density matrix on the Hilbert space
$L^2(\R^d)\otimes \mathcal H_2$. For every fixed $\x\in\R^n$
we can define $\langle \x|\rho_{12}|\x\rangle$ as an operator
on $\mathcal H_2$. This really means the partial trace
$\tr_1$ of $A_j^* \rho_{12} A_j$. Likewise we define
$\langle \kk|\rho_{12}|\kk\rangle$ to be $\tr_1 B_k^* \rho_{12} B_k$. We
see that $\langle \x|\rho_{12}|\x\rangle$ is an
operator-valued density in position space and $\langle 
\kk|\rho_{12}|\kk\rangle$ is the corresponding density in
momentum space. 

Now we apply the 2--space theorem and infer that
\begin{equation}
 \label{eq:fourier}
 H(1^A|2) + H(1^B|2) \geq S(1|2) \,,
\end{equation}
where
$$
H(1^A|2) = - \int_{\R^d} d^d\x \tr_2 \langle
\x|\rho_{12}|\x\rangle \ln \langle \x|\rho_{12}|\x\rangle -
S(\rho_2)
$$
and
$$
H(1^B|2) = - \int_{\R^d} d^d\kk \tr_2 \langle
\kk|\rho_{12}|\kk\rangle \ln \langle \kk|\rho_{12}|\kk\rangle -
S(\rho_2) \,.
$$
Here $c_1 = c_\infty = \sup_{\x,\kk} |e^{i2\pi \kk\cdot \x}| =1$
and $\ln c_1=\ln c_\infty =0$. This is a generalization of the
uncertainty principle in \cite{FrLi}, which is what
\eqref{eq:fourier} reduces to when
$\rho_{12}=\rho_1\otimes\rho_2$ is a product state. The
3--space theorem and the 2$\frac12$--space theorem obviously
generalize in a similar way for the Fourier transform.

This example of a continuum version of our theorems has the
obvious generalization to positive operator-valued
measures (POVMs). The interested reader can work this out
for him/herself, but we mention here one further specific
generalization. Let us take $\mathcal H_1=L^2(X,d\x)$, where
$X$ is some configuration space (e.g., $X=\R^d$ or $X=$ a
torus or $X=$ a lattice) and $L^2(X,d\x)$ are the square
integrable functions on $X$ with respect to some measure
$d\x$. The measurements $A_j^*A_j$ are again given by rank one
projections corresponding to the functions $\delta_{\x'}(\x)$
for some $\x'$ in $X$. Now let $\mathcal H_1'$ be a second
Hilbert space of the form $L^2(K,d\kk)$ and let $\mathcal U$
be a unitary from $\mathcal H_1$ to $\mathcal H_1'$, which
is given by a kernel $\mathcal U(\kk,\x)$. For each $\kk\in K$,
we can think of $\mathcal U(\kk,\x)$ as a function of $\x$ and
we define the $B_k^*B_k$'s to be the rank-one projections onto
these functions. In this way we obtain, as before, operators
$\langle \x|\rho_{12}|\x\rangle$ and $\langle
\kk|\rho_{12}|\kk\rangle$ on $\mathcal H_2$. Now $H(1^A|2)$ and $H(1^B|2)$ are
defined as in the Fourier transform case, except that the
integration is over the sets $X$ and $K$, respectively.

\textbf{Generalized Fourier transform theorem.}---
\eqref{eq:2sp} is valid with
$$
c_1 = \sup_{\kk,\x} |\mathcal U(\kk,\x)| \,.
$$

An examples in which the classical entropies are simultaneously discrete and continuous is the following. Suppose $X=\Z$, i.e., the integers, like the sites in a tight binding model. An $L^2$ function is the wave function of an itinerant electron. The second space $\mathcal H_1'$ is the square integrable functions on $K=(-1/2,1/2)$, the Brillouin zone. In this case the delta-functions, $\delta_{x x'}$, on $X$ are legitimate Kronecker deltas, and the (normalized) plane waves on $X$, parametrized by $k\in K$, are $e^{2\pi i kx}$. The unitary here is $\mathcal U(k,x)= e^{2\pi i kx}$ and therefore $c_1=1$. Since $dx$ is counting measure on $\Z$ the expression of $H(1^A|2)$ is a sum $\sum_{x\in \Z}$, whereas $dk$ is ordinary Lebesgue measure on $(-1/2,1/2)$ and $H(1^B|2)$ is an integral $\int_K dk$.

In conclusion, we have shown several things. (1) The entropy inequalities in \cite{BeChCoReRe,CoYuGhGr} can be proved in a few lines essentially by imitating the original proof of strong subadditivity of entropy. (2) We have carried at least one of these inequalities forward by utilizing the trace norm $c_1$ instead of the operator norm $c_\infty$. (3) We have shown how these inequalities, suitably interpreted, extend the 2--space entropy uncertainty principle to continuous bases such as position and momentum.

\smallskip
We thank Zhihao Ma for helpful correspondence. U.S.~National Science Foundation grants PHY-1068285 (R.F.) and PHY-0965859 (E.L.) are acknowledged.


\bibliographystyle{amsalpha}

\end{document}